\documentclass[pra,english,notitlepage,twocolumn,superscriptaddress]{revtex4-1}
\usepackage{amssymb}
\usepackage{hyperref}
\usepackage{graphicx}
\usepackage{amsmath}
\usepackage{color}
\usepackage{float}
\usepackage{xcolor}
\pagecolor{white}

\makeatletter
\renewcommand*\env@matrix[1][\arraystretch]{%
  \edef\arraystretch{#1}%
  \hskip -\arraycolsep
  \let\@ifnextchar\new@ifnextchar
  \array{*\c@MaxMatrixCols c}}
\makeatother

\newcommand{\COMMENTED}[1]{}

\begin{document}

\title{Topological Superconductivity in a two-dimensional Weyl SSH model}
\author{Peter Rosenberg}
\affiliation{Département de Physique \& Institut Quantique, Université de Sherbrooke, Québec, Canada J1K 2R1}
\author{Efstratios Manousakis}
\affiliation{National High Magnetic Field Laboratory and Department of Physics, Florida State University, Tallahassee, Florida 32306, USA}
\affiliation{Department of Physics, National and Kapodistrian University of Athens, Panepistimioupolis, Zografos, 157 84 Athens, Greece}

\begin{abstract} 
  We study the emergence of topological superconductivity in a two-dimensional (2D) Weyl system, composed of 
  stacked Su-Schrieffer-Heeger (SSH) chains. A previous analysis of the model \cite{2DWeyl_AFQMC}, showed that the addition 
  of an attractive Hubbard interaction between spinful electrons leads to a superconducting state that has an intricate pairing structure, 
  but is topologically trivial. Here we consider a pairing
  interaction that couples spinless fermions on opposite sublattices within the same unit cell. We observe that this physically motivated,
  momentum-independent pairing interaction induces a topological superconducting state, characterized by a gap function with a non-trivial phase, 
  as well as Majorana and Fermi arc edge states along the edge perpendicular to the direction of the SSH dimerization. In addition,
  we observe a transition as a function of pairing interaction strength and chemical potential, indicated by a change in the sign of the topological 
  charge carried by each of the four Bogoliubov-Weyl nodes.
\end{abstract}

\maketitle

\section{Introduction}

The discovery of topological materials marked the beginning of a new era in condensed matter physics, centered on the role of topology in condensed matter systems \cite{RevModPhys.82.3045,bernevig}.
As the field has developed, many models have been constructed to describe these novel materials, but one of the earliest and most enduring is the Su-Schrieffer-Heeger
(SSH) model \cite{SSH1979}. The model, initially conceived to explain the emergence of solitons in quasi-one-dimensional materials like polyacetylene, is based on a simple
picture of one-dimensional chains with staggered hopping strengths \cite{Rice1982}. Despite this apparent simplicity, the model
has helped introduce various 
novel topological phenomena \cite{PhysRevLett.62.2747}, including charge fractionalization, solitons, edge states, and topological transitions \cite{SSH_RMP,Rice1982}.
The model has also been emulated using cold atoms in optical lattices, where the Zak phase \cite{PhysRevLett.62.2747} has been measured \cite{Atala2013} and
the so-called Thouless charge pumping \cite{PhysRevB.27.6083} has been realized as topological charge pumping \cite{PhysRevLett.111.026802,Nakajima2016,Lohse2016,Leder2016}.

Several variations of the SSH model have been developed and studied since its introduction, for instance
models that include longer range hopping \cite{PhysRevX.8.031045,PhysRevB.99.035146,Li2014} or
a two-leg ladder of SSH chains, known as the Creutz model \cite{PhysRevB.96.035139,PhysRevX.7.031057},
as well as other 2D extensions \cite{Tam2022}.
In two previous publications we studied superconductivity in a model consisting of an infinite array of SSH chains \cite{2DWeyl_AFQMC,3DChains}. 
In particular, we focused on a parameter regime in which the normal state of the system is a Weyl semimetal.

The observation of superconductivity in topological materials, including Dirac and Weyl semimetals \cite{Leng2017,Guguchia2017,MoTe2Rhodes_PRB2017,EnhancedTcMoTe2Rhodes_2019},
has opened a new avenue in the field of topological condensed matter, with a focus on the interplay of superconductivity and topology. Topological superconductors, characterized
by the presence of Majorana fermions, are believed to offer an ideal platform for quantum computing \cite{PhysRevLett.86.268,KITAEV20032,Alicea2012,PhysRevX.6.031016,PhysRevB.95.235305} and information, which has motivated a concentrated effort to 
characterize these states and discover their underlying mechanisms.

One well-studied class of topological superconductors comprises systems with momentum-dependent pairing interactions,
for instance with $p$-wave or $d$-wave symmetry \cite{Read2000,Ueno2013}. These systems are believed to support Majorana fermions with properties that make them well-suited for applications in quantum computing \cite{PhysRevLett.86.268,Nayak2008}. Yet, it has proven difficult to discover or design materials with
intrinsic $p$-wave pairing, that might be used in quantum computing applications. In the context of cold atoms, proposals exist to generate $p$-wave superfluids from $s$-wave interactions \cite{Zhang2008}. Recent work on superconductivity in Weyl semimetals may however provide an alternative route towards
real material realizations of topological superconductivity. These studies note that topological superconducting states can be induced in Weyl systems subject to conventional $s$-wave pairing 
interactions \cite{Balents_2012,Burkov_2015,faraei_induced_2019}. In the absence of interactions, these models describe Weyl systems
characterized by a set of nodal points carrying topological charge. Upon the introduction of interactions each node is split into a pair of Bogoliubov-Weyl 
nodes with zero-energy Majorana states at the boundaries.

Here we treat a lattice fermion model, which is a straightforward extension of the SSH model.
The non-interacting component of the Hamiltonian describes a set of 1D SSH chains that are coupled
via an inter-chain hopping term to form a 2D lattice. The model supports a Weyl semimetal state, with
a single pair of nodes. The topological properties of this model, as well as its intricate pairing behavior 
in the presence of an attractive Hubbard interaction, have been studied previously using a combined
mean-field theory and auxiliary-field quantum Monte Carlo approach \cite{2DWeyl_AFQMC}.

In the present work, we show that the presence of a momentum-independent inter-orbital pairing term coupling spinless fermions
in the same unit cell yields a topological superconducting state with a gap function that has emergent $p$-wave character in the vicinity
of the nodal points. This state is characterized by the presence of both zero-energy Majorana edge states and Bogoliubov-Weyl Fermi arc states. 
Our model provides a clear example of the emergence of topological superconductivity from a momentum-independent pairing term in a system 
with an underlying topological band structure, as well as an indication of the essential ingredients to realize topological superconducting states 
from momentum-independent interactions between bare particles.

\section{Model}
\label{sec:latt_ham}

We consider the following lattice Hamiltonian,

\begin{eqnarray}
  \hat{H}_0&=&-\sum_{n,m}vc^{(\textmd{A})\dagger}_{n,m}c^{(\textmd{B})}_{n,m}+wc^{(\textmd{A})\dagger}_{n,m}c^{(\textmd{B})}_{{n-1,m}}+\textrm{h.c.}\nonumber\\
    &&-\sum_{n,m}t_dc^{(\textmd{B})\dagger}_{{n,m}}
    c^{(\textmd{A})}_{{n,m\pm1}}
    +t_dc^{(\textmd{B})\dagger}_{{n,m}}
    c^{(\textmd{A})}_{{n+1,m\pm1}}+\textrm{h.c.}\nonumber\\
    &&+\sum_{n,m,\alpha}(\varepsilon^\alpha-\mu) n^{(\alpha)}_{n,m}
    \label{eq:H0}
\end{eqnarray}
where the operator $c^{(\textmd{A})\dagger}_{{n,m}}$ creates an electron on the A site of the unit cell at $\mathbf{R}_{nm}=na\,\hat{x}+mb\,\hat{y}$, where $a$ and $b$ are lattice spacings, and $n^{(\alpha)}_{n,m}$ is the number operator for sublattice $\alpha=\textmd{A,B}$ in the unit cell at $\mathbf{R}_{nm}$. Each unit cell contains one A site and one B site. The intra- and inter-unit-cell hopping strengths in the $\hat{x}$-direction are given by $v$ and $w$, respectively, and the diagonal hopping strength is given by $t_d$. We include an on-site potential term for each sublattice, $\varepsilon^{\textmd{A,B}}$, and a chemical potential, $\mu$. The first line of the Hamiltonian above corresponds to the standard 1D SSH model. These 1D SSH chains are coupled by the diagonal hopping included on the second line, which serves to extend the model to 2D. In all the results that follow we take $v=0.6$, $w=1.2$, and $t_d=0.9$.

The Hamiltonian in Eq.~(\ref{eq:H0}) can be viewed as a two-dimensional extension 
of the SSH hamiltonian. We choose to present the model as a set of stacked SSH 
chains for the purposes of conceptualization, but such a model can arise in other quasi-2D 
materials as a result of structural distortions, for example a Peierls or
Jahn-Teller distortion in one of the crystallographic directions,
causing a transition from a tetragonal to an orthorhombic phase.
In fact, such systems offer a better platform to realize the Weyl behavior
discussed here, because its presence requires the three hopping matrix
elememts, $v$, $w$ and $t_d$ be of comparable magnitude.

In momentum space the Hamiltonian can be written: 

\begin{equation}
\mathcal{H}_0(\mathbf{k})=-\mathbf{h}(\mathbf{k})\cdot\boldsymbol{\sigma}+\bar{\varepsilon}-\mu,
\end{equation}
where $\boldsymbol{\sigma}=(\sigma_x,\sigma_y,\sigma_z)$ is a vector of the Pauli matrices in the sublattice (pseudospin) basis, 
with $\bar{\varepsilon}=(\varepsilon^\textmd{A}+\varepsilon^\textmd{B})/2$, and
 $\delta\varepsilon = (\varepsilon^B-\varepsilon^A)/2$. When $\varepsilon^{\textmd{A}}\neq\varepsilon^{\textmd{B}}$
inversion symmetry is broken.
The vector $\mathbf{h}(\mathbf{k})$ has the form,
\begin{equation}\mathbf{h}(\mathbf{k})=
\begin{pmatrix}
v_1+w_1\cos(k_xa) \\
w_1\sin(k_xa) \\
\delta \varepsilon
\end{pmatrix},
\label{eq:hk}
\end{equation}
with $v_1=v+2t_d\cos(k_yb)$, and $w_1=w+2t_d\cos(k_yb)$. This model has two pseudo-helicity bands given by,
\begin{equation}
E^\pm(\mathbf{k}) = \tilde{\mu}\pm \vert \mathbf{h}(\mathbf{k}) \vert,
\label{eq:E0_pm}
\end{equation}
where $\tilde{\mu}=\bar{\varepsilon}-\mu$. The corresponding creation operators for states in these bands, $\chi^{\dagger(\pm)}_{\mathbf{k}}$, are related to the bare fermion operators
according to,
\begin{multline}
\begin{pmatrix}
\chi^{\dagger(-)}_{\mathbf{k}} & \chi^{\dagger(+)}_{\mathbf{k}}
\end{pmatrix}=
\begin{pmatrix}
c^{\dagger(\textmd{A})}_{\mathbf{k}} & c^{\dagger(\textmd{B})}_{\mathbf{k}}
\end{pmatrix}\times\\
\frac{1}{\sqrt{2}}
\begin{pmatrix}
\sqrt{1-\cos\phi_\mathbf{k}}\,e^{-i\theta_\mathbf{k}/2} && \sqrt{1+\cos\phi_\mathbf{k}}\,e^{-i\theta_\mathbf{k}/2} \\
-\sqrt{1+\cos\phi_\mathbf{k}}\,e^{i\theta_\mathbf{k}/2} && \sqrt{1-\cos\phi_\mathbf{k}}\,e^{i\theta_\mathbf{k}/2}
\end{pmatrix},
\label{eq:helicity_basis}
\end{multline}
where we have parametrized $\mathbf{h}(\mathbf{k})$ as $\mathbf{h}(\mathbf{k})=\vert \mathbf{h}(\mathbf{k}) \vert (\cos\theta_\mathbf{k}\sin\phi_\mathbf{k},
\sin\theta_\mathbf{k}\sin\phi_\mathbf{k},\cos\phi_\mathbf{k})$, with $\theta_\mathbf{k}\equiv\tan^{-1}(\mathbf{h}_y/\mathbf{h}_x)$, and 
$\phi_\mathbf{k}\equiv\tan^{-1}\left(\sqrt{\mathbf{h}_x^2+\mathbf{h}_y^2}/\mathbf{h}_z\right)$.

This band structure has two nodal points of opposite chirality, whose locations are given by the solution of
$E^+(\mathbf{k}) = E^-(\mathbf{k})$, which occurs at $k^\pm_N=(0,\pm k_y^N)$, with $k_y^N=\cos^{-1}(-(v+w)/4t_d)$.
Expanding $\mathbf{h}(\mathbf{k})$ in the vicinity of these nodal momenta (retaining only terms linear in $\mathbf{k}$), we obtain,
\begin{equation}
\mathbf{h}^\text{linear}_\mathbf{k}=
\begin{pmatrix}[1.5]
\pm 4t_d\sin(k^N_y)(k^0_y-k^N_y) \\
\frac{1}{2}(w-v)(k^0_x-k^N_x) \\
\delta\varepsilon
\end{pmatrix}.
\label{eq:h_linear} 
\end{equation}
The linear behavior of the dispersion near these nodal points helps confirm their Weyl character.

A topological characterization of this model can be found in Ref.~\cite{2DWeyl_AFQMC}, which also contains an investigation
of the effects of an attractive Hubbard interaction using a combination of mean-field theory and auxiliary-field quantum Monte Carlo calculations.
This study revealed that on-site interactions between like pseudo-spins leads to a superconducting phase with an intricate pairing structure, though
the Fermi arcs that are present in the normal state do not persist in the superconducting state, which is topologically trivial.

As noted above, in the context of Weyl systems, it has been shown that topological superconducting states 
can emerge from conventional $s$-wave pairing interactions \cite{Balents_2012,Burkov_2015,faraei_induced_2019}.
These models describe topological semimetals in the normal state, characterized by the presence of nodal points that
carry topological charge. In the superconducting state, each node splits into a pair of Bogoliubov-Weyl nodes, and 
zero-energy Majorana states appear at the boundaries.

\begin{figure*}[ht]
    \begin{center}
           \includegraphics[width=0.95\textwidth]{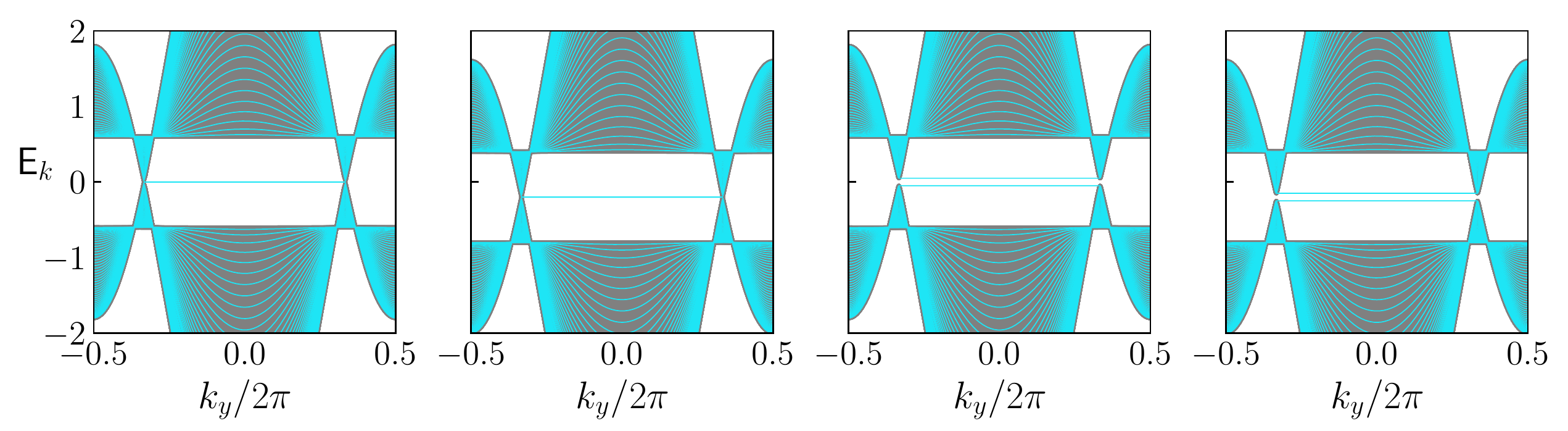}
    \end{center}
    \caption{           \label{fig:ek0_ek_NI}
Edge Spectrum for non-interacting system. The color curves show the results from a slab calculation, while the bulk is shown in gray. From left to right, $(\mu, \delta\varepsilon)=(0,0), (0.2, 0), (0, 0.1), (0.2,0.1)$. 
}
\end{figure*}

\begin{figure*}[ht]
    \begin{center}
           \includegraphics[width=0.95\textwidth]{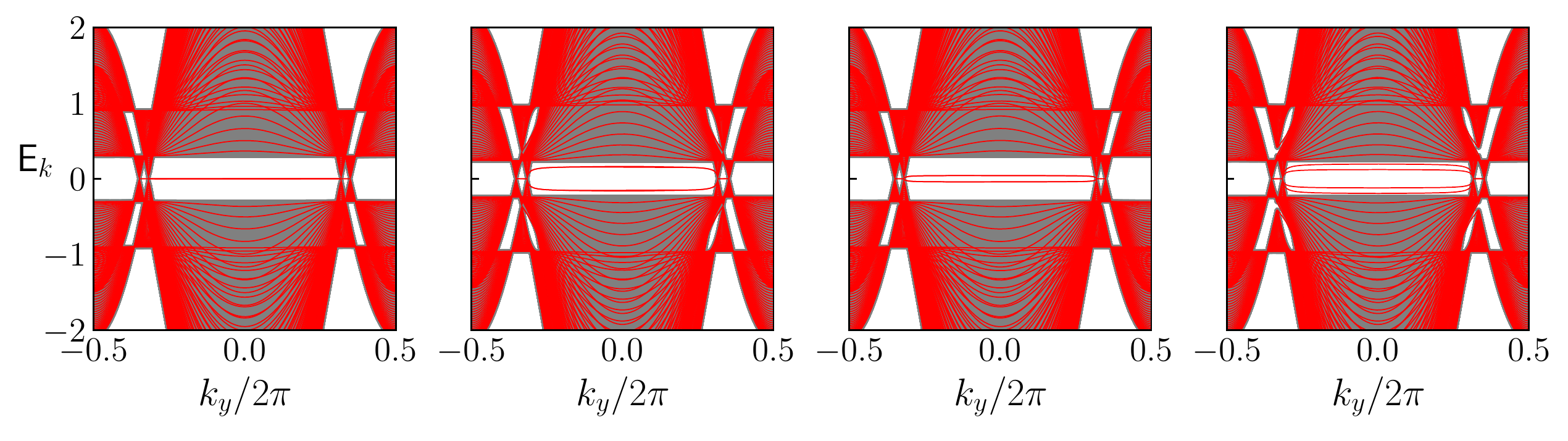}
    \end{center}
    \caption{           \label{fig:ek0_ek_INT}
Edge Spectrum for interacting system with $\Delta_0 = 0.3$. From left to right, $(\mu, \delta\varepsilon)=(0,0), (0.2, 0), (0, 0.1), (0.2,0.1)$. 
}
\end{figure*}

The model defined in Eq.~(\ref{eq:H0}) describes a topological semimetal in the normal state.
Building on this model, we consider an interaction term that couples particles on 
opposite sublattices within the same unit cell. We observe that this interaction induces a topological superconducting state that supports both 
Bogoliubov-Weyl Fermi arcs and Majorana edge modes, without requiring more complicated momentum-dependent pairing terms, for instance 
of $p$-wave or $d$-wave form. Our observation of topological superconducting states in this model provides
a simple description of the emergence of topological superconductivity from normal-state topological semimetals, and suggests that 
standard $s$-wave pairing interactions are sufficient to realize these states.

After adding this pairing term to our model, the total Hamiltonian can be written in the pseudo-spin basis as,
\begin{equation}
\hat{H} = \sum_{\mathbf{k}\alpha\beta} c^{\dagger(\alpha)}_{\mathbf{k}}\mathcal{H}_0(\mathbf{k})_{\alpha\beta} c^{(\beta)}_{\mathbf{k}}
+\sum_{\mathbf{k}}\Delta_0 c^{\dagger(\textmd{A})}_\mathbf{k}c^{\dagger(\textmd{B})}_\mathbf{-k}+\textmd{h.c.},
\label{eq:H_pseudospin_basis}
\end{equation}
where the pairing amplitude $\Delta_0$ is a momentum-independent constant.

We choose this pairing term based on its natural physical motivation. It is quite possible
that this simple type of pairing interaction could arise in a quasi-2D system, such  
as the one described by our tight-binding Hamiltonian, Eq.~\ref{eq:H0}, in its normal state.  
This Hamiltonian possesses a $(\pi,0)$ charge-density-wave (CDW) order
in the normal state. It has been well-established in several quantum many-body systems,
for instance the cuprate superconductors, that the CDW order
competes with the superconducting order. While these two types of order can co-exist, in some cases
partially or completely removing the CDW order opens up the superconductivity channel.
In our model, the onset of the CDW order is an indication of an effective attraction between fermions
that leads to increased density along alternating bonds between the A and B sublattices.
It is reasonable to assume that this effective attraction will manifest itself in the superconducting
channel if the CDW order is somehow suppressed, for instance by doping, which would likely lead to
a pairing interaction between fermions on neighboring A and B atoms. This simple, though physically 
grounded picture suggests that the pairing term we consider might be naturally realized in a wide class of
quasi-2D materials.

In the pseudo-helicity basis the Hamiltonian in Eq.~(\ref{eq:H_pseudospin_basis}) can be written,
\begin{equation}
\hat{H} = \sum_{\mathbf{k,\alpha=\pm}} E^{\alpha}(\mathbf{k})\chi^{\dagger(\alpha)}_\mathbf{k}\chi^{(\alpha)}_\mathbf{k}
+\sum_{\mathbf{k}}\Delta^\chi(\mathbf{k}) \chi^{\dagger(\textmd{-})}_\mathbf{k}\chi^{\dagger(\textmd{+})}_\mathbf{-k}+\textmd{h.c.},
\label{eq:H_pseudohelicity_basis}
\end{equation}
where $E^{\pm}(\mathbf{k})$ is defined in Eq.~(\ref{eq:E0_pm}) and $\Delta^\chi(\mathbf{k})=\Delta_0(1+\cos\phi_\mathbf{k})e^{-i\theta_\mathbf{k}}$. 
We note that the gap function $\Delta^\chi(\mathbf{k})$ has a momentum-dependent phase, whose behavior is connected to the topological character of the non-interacting band
structure. Making use of Eq.~(\ref{eq:h_linear}) we find that in the vicinity of the nodal points the gap function $\Delta^\chi(\mathbf{k})\propto \alpha(k^0_y-k^N_y)+i\beta(k^0_x-k^N_x)$,
where $\alpha,\beta$ are real coefficients. This $p$-wave structure naturally emerges from the simple momentum-independent pairing term we have introduced into the model.
In other words, the gap function acquires a non-trivial phase from the normal state band structure. This is a similar observation to previous studies on pairing in
spinless models that possess topological band structures in the normal state~\cite{Murakami_2003,Haldane_2018}, though unlike these studies our lattice model 
preserves time reversal symmetry. 

\begin{figure}[ht!]
    \begin{center}
           \includegraphics[width=0.9\columnwidth]{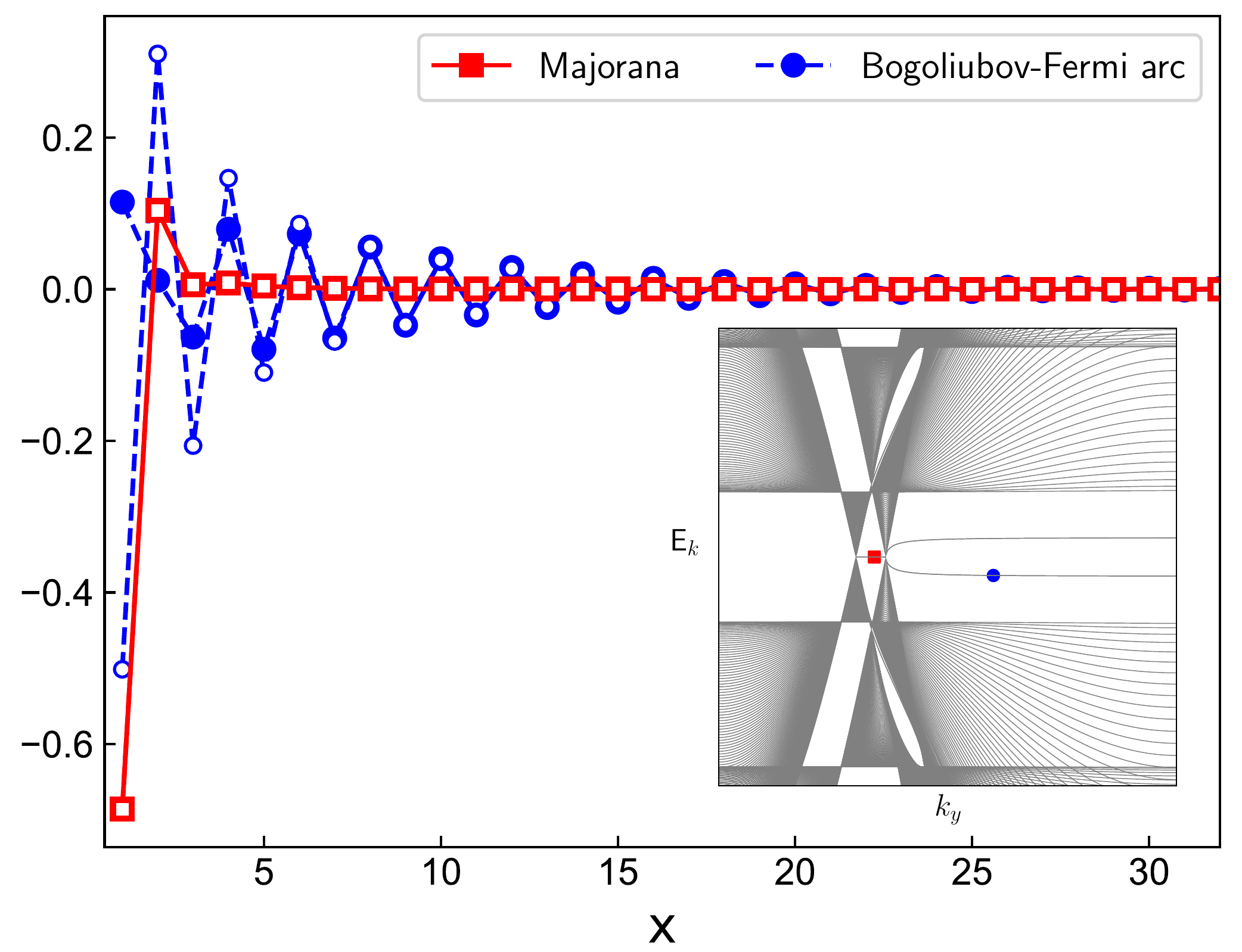}
    \end{center}
    \caption{           \label{fig:wf_amps}
Wavefunction amplitudes versus $x$. We plot the components $a_{i,k_y}$ (filled symbols) and $(c_{i,-k_y})^\ast$ (empty symbols) of the eigenvectors for a Majorana state and a Bogoliubov-Fermi arc state. Per the condition in Eq.~(\ref{eq:Maj_cond}) we
classify the state with $a_{i,k_y}=(c_{i,-ky})^\ast$ as a Majorana state, and the state with $a_{i,k_y}\neq(c_{i,-ky})^\ast$ as a Bogoliubov-Fermi arc state.
The inset shows the edge spectrum with the symbols indicating the momentum corresponding to the Majorana (red square) and Bogoliubov-Fermi arc (blue circle) states shown in the main panel. We take $\Delta_0=0.3$ and $\mu=0.1$.
}
\end{figure}

The full Hamiltonian can be diagonalized exactly, yielding eigenvalues,
\begin{align}
E^1_\pm(\mathbf{k})&=\pm\sqrt{\eta^2+2\vert\zeta\vert}\notag\\
E^2_\pm(\mathbf{k})&=\pm\sqrt{\eta^2-2\vert\zeta\vert},
\end{align}
where we have defined,
\begin{align}
\eta^2 &= \tilde{\mu}^2-\delta\varepsilon^2-\vert h_\mathbf{k} \vert^2 + \vert\Delta\vert^2 \\
\zeta^2 &=(\vert h_\mathbf{k}\vert^2+{\delta\varepsilon}^2)\tilde{\mu}^2+({h^x_\mathbf{k}}^2+\delta\varepsilon^2)\vert\Delta \vert^2.
\end{align}
There are 4 nodes occurring where $E^2_+(\mathbf{k})=E^2_-(\mathbf{k})=0$.
This leads to the condition,
\begin{equation}
\eta^2=2\vert\zeta\vert,
\end{equation}
which has four solutions at $k_x=0$,
\begin{equation}
\pm k^{N,\pm}_y = \cos^{-1}\left[\frac{\pm \sqrt{\tilde{\mu}^2-\delta\varepsilon^2+\vert\Delta\vert^2} -(v+w)}{4t_d}\right].
\label{eq:kN}
\end{equation}
These correspond to the two original nodal points at $(0,\pm \cos^{-1}(-(v+w)/4t_d))$, which are now split into four nodal points, with
the pairs of nodes centered on $(0,\pm \cos^{-1}(-(v+w)/4t_d))$, but shifted by the finite $\tilde{\mu}$, $\delta\varepsilon$ and $\Delta$. In the case of $\tilde{\mu}=\delta\varepsilon=\Delta=0$,
we recover the non-interacting result. We note that while the non-interacting spectrum is gapped for $\delta\varepsilon\neq 0$, this gap can be closed with the addition of a
pairing interaction.

\section{Topological Superconductivity}

    \begin{center}
\begin{figure*}[ht]
           \includegraphics[width=\textwidth]{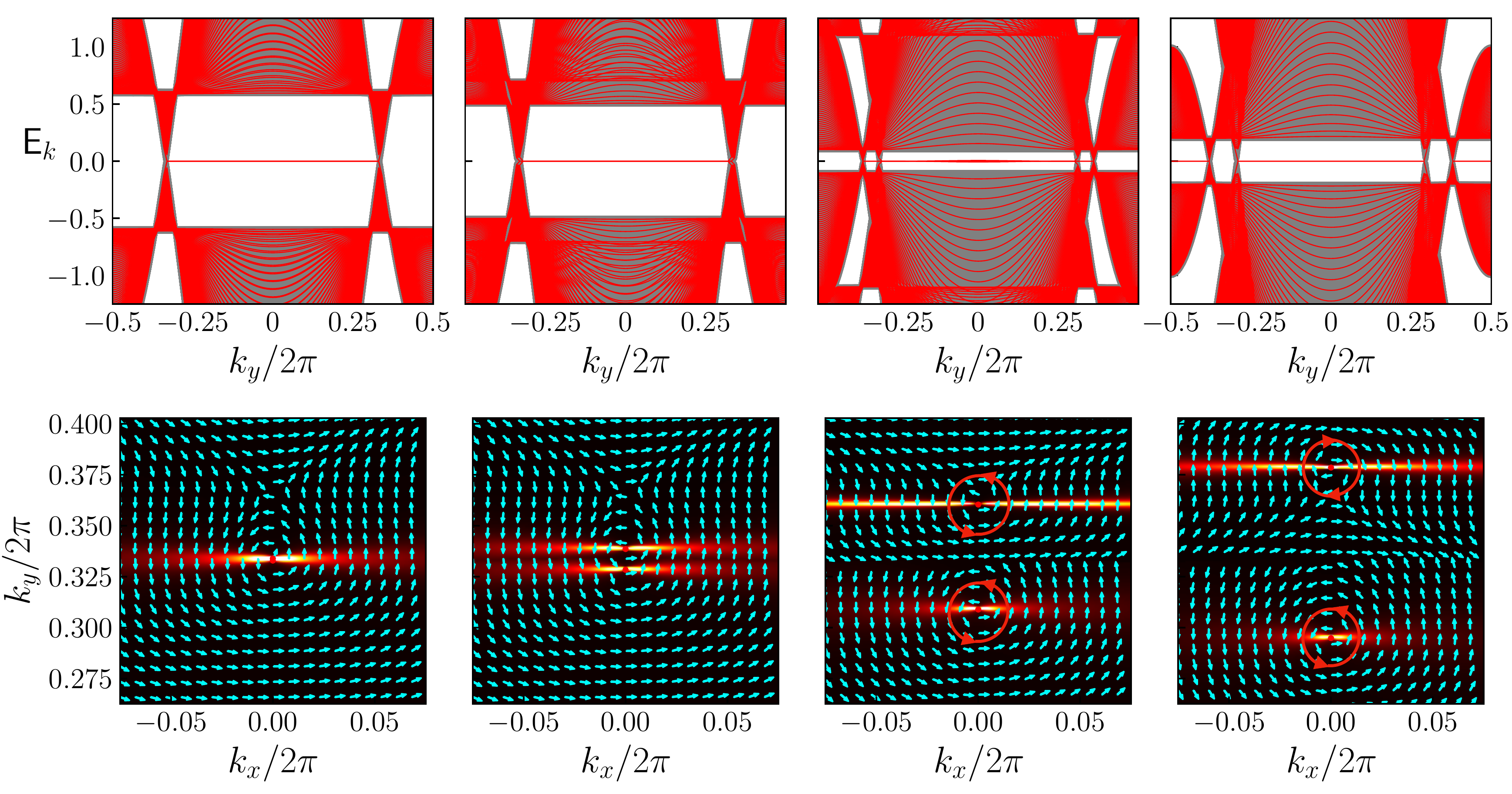}
    \caption{           \label{fig:ek_int_vs_delta}
Edge Spectrum and Berry connection vector versus $\Delta$ at $\mu=0.0$. (a) $\Delta=0.01$, (b) $\Delta=0.1$, (c) $\Delta=0.5$, (d) $\Delta=0.8$. The top
row shows the edge spectrum (red curves), the bulk region is shown in gray. The bottom row shows the Berry connection vector around the pair of nodes in
the upper BZ, with the direction given by the cyan arrows and the magnitude given by the color density plot. The red circles with arrows are a 
guide to the eye showing the circulation pattern of the Berry connection vector around either node (indicated by the small red dot) 
in the upper half of the BZ, before and after the transition.
}
\end{figure*}
    \end{center}

The analysis above shows that the addition of a pairing term coupling fermions on different sublattices within the same
unit cell leads to the splitting of each original nodal point into a pair of Bogoliubov-Weyl nodes. In order to characterize 
the topological nature of this state we proceed by computing the edge spectrum and the Berry connection vector, which
provide clear evidence of Majorana modes and a topological transition as a function of the pairing strength, $\Delta_0$, and
the chemical potential, $\mu$. 

\subsection{Edge spectrum and Majorana states}
To obtain the edge spectrum for the system, we diagonalize the Hamiltonian in a slab geometry (with finite $x$ dimension). The edge spectrum reveals the
presence of both Majorana states at zero energy and Fermi arc states. We first present an example of the edge spectrum for
the non-interacting model.
The non-interacting system remains gapless for finite chemical potential (Fig.~(\ref{fig:ek0_ek_NI})(b)), which simply shifts the Fermi level.
With finite $\delta\varepsilon$ a gap opens, separating the pair of Fermi arc states (Fig.~\ref{fig:ek0_ek_NI}(c)). When $\mu$ and
$\delta\varepsilon$ are both finite the Fermi level shifts and a gap opens (Fig.~\ref{fig:ek0_ek_NI}(d)). 

In the interacting system the pair of Weyl nodes is split into two
pairs of Bogoliubov-Weyl nodes. There are four Fermi arc states passing through the origin that connect the inner nodes. These four states are degenerate at 
$\mu=\delta\varepsilon=0$ (Fig.~\ref{fig:ek0_ek_INT}(a)). The degeneracy is partially lifted by finite $\mu$ (Fig.~\ref{fig:ek0_ek_INT}(b)), which splits the states 
into two pairs related by particle-hole symmetry, and also by finite $\delta\varepsilon$ (Fig.~\ref{fig:ek0_ek_INT}(c)). With finite $\mu$ and finite $\delta\varepsilon$ the 
degeneracy is fully lifted (Fig.~\ref{fig:ek0_ek_INT}(d)). In addition to the Fermi arc states there is a pair of Majorana states at zero energy connecting the inner node near
$k^+_N$ to the outer node near $k^+_N$ and similarly for the pair of nodes around $k^-_N$. 

In order to characterize the topological nature of the edge states we examine the components of their wave functions. 
Majorana states are defined by the property of their creation and annihilation operators, 
$\gamma_\mathbf{k}^\dagger=\gamma_\mathbf{-k}$. 
For the case of finite $x$ dimension, the Hamiltonian can be diagonalized by Bogoliubov operators of the
general form,
\begin{equation}
\gamma^{n\dagger}_{k_y} = \sum_i a^n_{i,k_y} \chi^{\dagger(+)}_{i,k_y}+b^n_{i,k_y} \chi^{\dagger(-)}_{i,k_y} +c^n_{i,k_y} \chi^{(+)}_{i,-k_y} + d^n_{i,k_y} \chi^{(-)}_{i,-k_y},
\end{equation}
where $n$ labels the eigenstate, and $i$ labels the layer in the finite direction.
The Majorana condition $\gamma_\mathbf{k}^\dagger=\gamma_\mathbf{-k}$, leads to the requirement (dropping the label $n$),
\begin{align}
\sum_i &a_{i,k_y} \chi^{\dagger(+)}_{i,k_y}+b_{i,k_y} \chi^{\dagger(-)}_{i,k_y} +c_{i,k_y} \chi^{(+)}_{i,-k_y} + d_{i,k_y} \chi^{(-)}_{i,-k_y} = \notag\\
\sum_j &(a_{j,-k_y})^\ast \chi^{(+)}_{j,-k_y}+(b_{j,-k_y})^\ast \chi^{(-)}_{j,-k_y} +(c_{j,-k_y})^\ast \chi^{\dagger(+)}_{j,k_y} \notag \\
&+ (d_{j,-k_y})^\ast \chi^{\dagger(-)}_{j,k_y}, 
\end{align}
which implies for a Majorana state (up to an arbitrary phase),
\begin{align}
a_{i,k_y}&=(c_{i,-k_y})^\ast,\quad b_{i,k_y}=(d_{i,-k_y})^\ast \notag \\
c_{i,k_y}&=(a_{i,-k_y})^\ast, \quad d_{i,k_y}=(b_{i,-k_y})^\ast.
\label{eq:Maj_cond}
\end{align}

In Fig.~(\ref{fig:wf_amps}) we plot the amplitudes $a_{i,k_y}$, and $(c_{i,-k_y})^\ast$ of the wavefunction for two states,
both of which are localized on the edge of the system.
We identify one state as Majorana due to the relationship between its amplitudes, as described above (though not shown,
the remaining coefficients satisfy the same condition). The other
state shown is a Bogoliubov-Fermi arc state, which lies in the bulk gap but is not pinned to zero energy at finite chemical potential
and does not satisfy the Majorana condition. 

\subsection{Topological transition}

\begin{figure*}[ht!]
    \begin{center}
           \includegraphics[width=\textwidth]{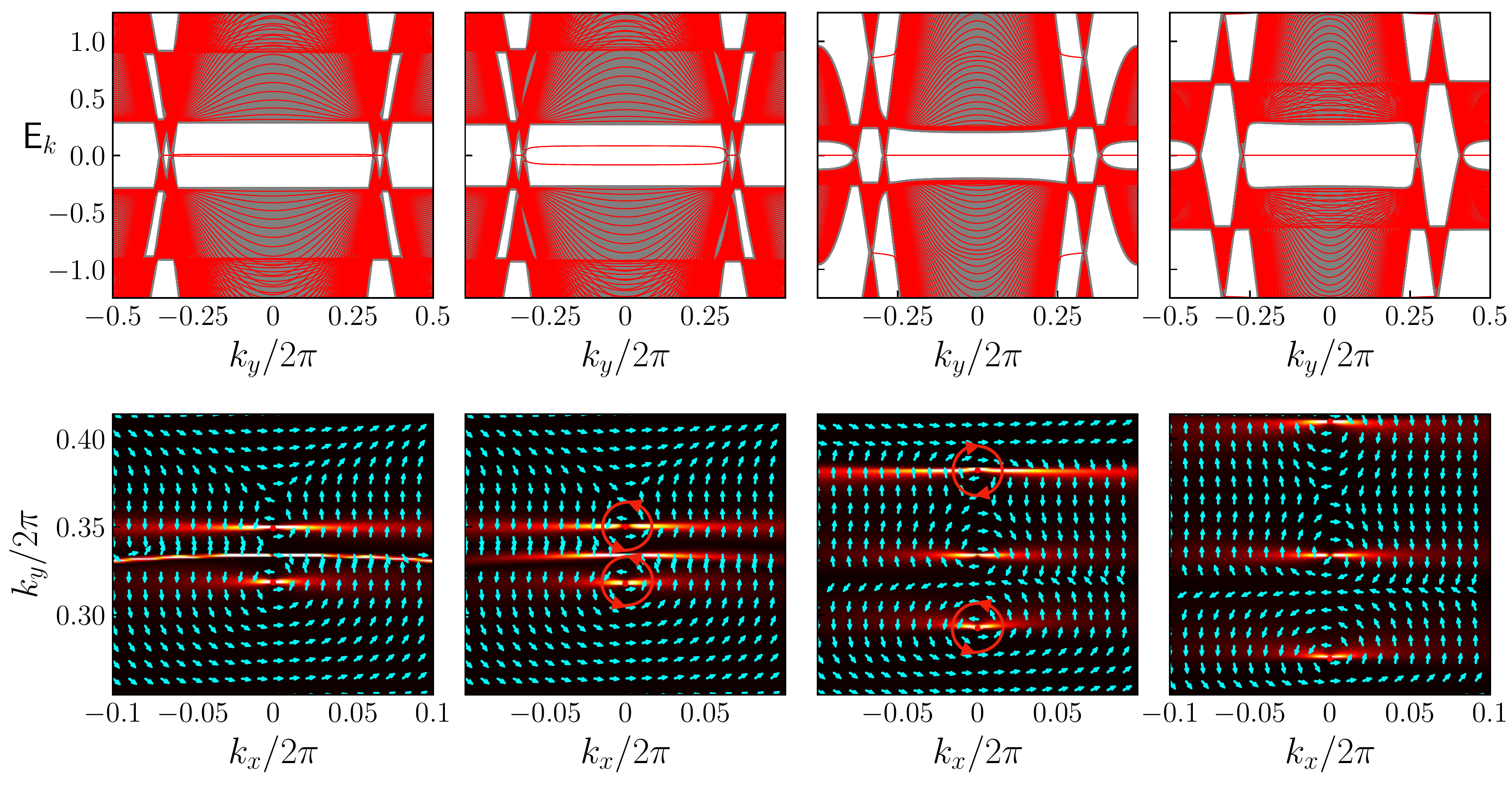}
    \end{center}
    \caption{           \label{fig:ek_int_vs_mu}
Edge Spectrum and Berry connection vector versus $\mu$ at $\Delta=0.3$. (a) $\mu=0.01$, (b) $\mu=0.1$, (c) $\mu=0.8$, (d) $\mu=1.2$.
The top row shows the edge spectrum (red curves), the bulk region is shown in gray. The bottom row shoes the Berry connection vector around the pair of nodes in
the upper BZ, with the direction given by the cyan arrows and the magnitude given by the color density plot. The red circles with arrows are a 
guide to the eye showing the circulation pattern of the Berry connection vector around either node (indicated by the small red dot) 
in the upper half of the BZ, before and after the transition.
}
\end{figure*}

Having confirmed the topological nature of the superconducting state that emerges in the interacting model,
which is characterized by the presence of Majorana states in the edge spectrum, we now
study the effect of chemical potential and pairing strength on this state. As we describe in the following, we observe
a topological transition as a function of pairing strength and chemical potential, marked by a qualitative change in the
edge spectrum and the behavior of the Berry connection vector.

In Fig.~\ref{fig:ek_int_vs_delta} we examine the edge spectrum and Berry connection vector versus $\Delta$ at fixed chemical potential. 
The Berry connection vector is calculated as $  {\bf A} \equiv - i \langle \psi^- | \nabla_{\bf k} | \psi^- \rangle$, where $\vert \psi^- \rangle$ is
the wavefunction corresponding to the lower Bogoliubov-Weyl band. This yields, 
\begin{eqnarray}
{\bf A} =
  {1 \over 2}  \nabla_{\bf k} \theta_{\bf k},
  \label{phase-eq}
\end{eqnarray}
where $\theta_\mathbf{k}$ is a parameter characterizing $\mathbf{h}(\mathbf{k})$ (Eq.~(\ref{eq:hk})), which defines the non-interacting Hamiltionian (Eq.~(\ref{eq:H0})), as defined above.

In the limit of $\Delta\rightarrow0$ (Fig.~\ref{fig:ek_int_vs_delta}(a)), the edge spectrum and the Berry connection vector resemble those of the non-interacting model, which is characterized by a Fermi arc state connecting nodes of opposite chirality. As described above, at finite $\Delta$ each of the original nodal points splits into a pair of Bogoliubov-Weyl nodes. At small $\Delta$ (Fig.~\ref{fig:ek_int_vs_delta}(b)-(c)) there are four Bogoliubov-Weyl Fermi arc states connecting the nodes closest to the origin. These states are degenerate at $\mu=\delta\varepsilon=0$. With increasing $\Delta$, at a fixed value of chemical potential, each pair of Bogoliubov-Weyl nodes begins to separate, and the Berry connnection
begins to deviate from the non-interacting result. Further increasing $\Delta$ leads to the emergence of four distinct sources of topological charge (two in each half of the Brillouin zone), as indicated by the Berry connection vector. Below a certain critical value of $\Delta$ these charges are of the same sign on a given half of the Brillouin zone, as indicated by the direction of circulation of the Berry connection vector around the node (Fig.~\ref{fig:ek_int_vs_delta}(c)).  Above this critical value of $\Delta$ we observe a transition to a state with topological charges of opposite sign on the same half of the Brillouin zone (Fig.~\ref{fig:ek_int_vs_delta}(d)), indicated by the opposite direction of circulation of the Berry connection around either node on the same half of the Brillouin zone. 

This transition is also reflected in the edge spectrum, by the closing and reopening of the bulk gap. At small to intermediate $\Delta$, below the critical value, we find a pair of Majorana states connecting each pair of Bogoliubov-Weyl nodes, in addition to four Bogoliubov-Weyl fermi arc states connecting the nodes closer to the origin of the Brillouin zone. After the gap closes and reopens we find four Majorana states. We observe a similar transition as a function of increasing $\mu$ for a fixed value of $\Delta$ (Fig.~\ref{fig:ek_int_vs_mu}).

\section{Conclusion}

In this work we have constructed a lattice model that extends the well-known SSH
model into two dimensions. In the absence of interactions, the model hosts a pair
of Weyl nodes in its band structure. The addition of a momentum independent pairing
interaction coupling fermions on different sublattices within the same unit cell splits
each node into a pair of Bogoliubov-Weyl nodes and induces a topological superconducting state, 
characterized by a pairing function with a non-trivial phase whose origin is connected to the 
topological band structure of the normal state.
The topological superconductor state supports both Majorana and Fermi arc states, whose
number and orientation is sensitive to the pairing strength and chemical potential, which
induce a topological transition marked by the closing and reopening of the bulk gap
accompanied by a qualitative change in the edge spectrum.

Given the potential applications of topological superconductors, for instance in the
field of quantum computing and information, a more complete understanding of the
origins of these states remains an important goal. Here we have presented a simple
model with a physically motivated pairing interaction, that serves as a clear illustration
of the emergence of a topological superconducting state from a Weyl semimetal system.
This description helps shed light on the connection between topological superconductivity and 
topological band structures and suggests compelling new avenues in the search for material
realizations of topological superconductivity.

\end{document}